# The Mechanism of Spin-Phonon Relaxation in Endohedral Metallofullene Single Molecule Magnets

Tanu Sharma[a], Rupesh Kumar Tiwari[a], Sourav Dey[a], Lorenzo A. Mariano[b], Alessandro Lunghi[b]* and Gopalan Rajaraman[a]*

- a- Department of Chemistry, Indian Institute of Technology Bombay, Mumbai, Maharashtra, 400076, India. Email: rajaraman@chem.iitb.ac.in.
- b- School of Physics and AMBER Research Centre, Trinity College, Dublin 2, Ireland. Email: lunghia@tcd.ie

**Abstract**

This study presents the first-ever investigation of spin-phonon coupling mechanisms in fullerene-based single-molecule magnets (SMMs) using ab initio CASSCF combined with DFT calculations. While lanthanide-based SMMs, particularly those with $Dy^{III}$ ions, are known for their impressive blocking temperatures and relaxation barriers, endohedral metallofullerene (EMFs) offer a unique platform for housing low-coordinated lanthanides within rigid carbon cages. We have explored the spin dynamics of in $DyScS@C_{82}$ exhibiting among the highest blocking temperature ($T_B$) reported. Through our computational analysis, we reveal that while the fullerene cage enhances crystal field splitting and provides structural stability without significantly contributing to spin-relaxation-driving low-energy phonons, the internal ionic motion emerges as the primary factor controlling spin relaxation and limiting blocking temperature. This computational investigation into the spin dynamics of EMF-based SMMs provides key insights into their magnetic behaviour for the first time and suggests potential strategies for improving their performance towards futuristic SMMs.

**Introduction**

Due to their very high blocking temperatures and large thermal barriers to relaxation, Dy$^{III}$-based single-molecule magnets (SMMs) are of great interest.[1-6] Among many challenges in taking these molecules to end-user applications, enhancing the barrier height for magnetization reversal ($U_{eff}$) and blocking temperature below which the magnetisation is fully frozen ($T_B$) are considered crucial. Coupling magnetic lanthanide ions to strong axial ligand fields is one of the most promising approaches to producing high-performing SMMs.[1, 3] This has produced SMMs with a barrier height of magnetisation reversal as large as 1500 cm$^{-1}$, achieving one of the aforementioned goals.[1, 2] Although such large barriers have been achieved so far the $T_B$ values remain modest at 80K.[1] While theoretical advances have propelled the understanding of $U_{eff}$ through various intuitive ligand design principles that have shaped the field of lanthanide SMMs over the past few decades, the comprehension of $T_B$—the most critical factor governing magnetic performance—remains largely unclear. This is evident from the fact that many molecules with high $U_{eff}$ display low $T_B$, and vice versa, underscoring the need for a molecular-level understanding beyond simple electron density considerations (such as prolate versus oblate shapes). To achieve the next breakthrough in enhancing $T_B$ values, it is essential to focus on this deeper molecular understanding. The $T_B$ values and the corresponding relaxation times are strongly linked to the spin-phonon relaxation mechanism, an area of significant recent research aimed at providing insights to improve $T_B$.[1, 2, 7-9]

The mechanism of spin-phonon relaxation has been explored so far in a handful of examples offering some guidelines to enhance the $T_B$ values.[8, 10, 11] These guidelines include: (i) achieving strong crystal-field splitting of the Kramers doublets,[12, 13] (ii) ensuring that the crystal field splitting exhibits minimal transverse anisotropy, which drives quantum tunnelling of magnetization (QTM),[14, 15] (iii) reducing the vibrational density of states at resonant frequencies,[14] (iv) minimizing low-energy vibrations,[16] (v) utilizing rigid ligands that can isolate intramolecular motions from low-energy acoustic vibrations[7, 17], and (vi) employing ligands with donor atoms whose local charge remains stable despite local vibrations.[8] Controlling these factors is very challenging, as crystal field parameters and reducing transverse anisotropy are correlated to the geometry and local point group of the molecules. Strong crystal field splitting and less transverse anisotropy can be achieved with low-coordinated lanthanides or with higher oxidation state,[18] but synthesizing such molecules is challenging, and even if they are synthesised with very bulky ligands, they often encounter other weak non-covalent interactions such as Ln…H-C agostic interactions that facilitate transverse anisotropy as demonstrated in several cases.[2, 19, 20] Moreover, traditional organometallic complexes often have ligands with loosely bound atoms, like hydrogen, whose vibrations also contribute to spin-phonon relaxation.[2] Synthesis of the rigid lanthanide low-coordinate molecule without atoms such as -H are extremely challenging and has not yet been achieved.

However, in the endohedral metallofullerenes (EMFs) class of lanthanide-based SMMs, these conditions can be met easily.[21-26] The EMF-based SMMs feature rigid cages that can stabilize low-coordinated lanthanides inside their cage structure.[26, 27] These compounds contain atoms or small clusters of atoms encapsulated by fullerenes, where cage shields the magnetic ion from the decoherence caused by external noise and hence stabilize atomic configurations that are not achievable in conventional molecules. These unique circumstances can produce very effective, precisely controlled SMMs. Although not all conditions are met, this type of molecule often satisfies conditions (i), (ii), (iv) (v) and (vi). This is demonstrated by the fact that lanthanide EMFs are reported as high-performing SMMs far more frequently than any other class of molecules (see Table S1). Despite these key advantages, the $T_B$ that has been reached with this class of molecule is relatively small compared to the best-in-class dysprosocenium class of molecules. This is linked to the spin-phonon relaxation mechanism, but a clear understanding of this process for this class of SMMs is still missing. Investigating the spin-phonon relaxation mechanism in these fullerene molecules could provide valuable insights, potentially leading to methods for suppressing or minimizing this effect.

Endohedral metallofullerenes can be broadly classified into two main categories. The first category consists of traditional metallofullerenes, where only metal atoms are encapsulated within the fullerene structure. In contrast, the second category, known as cluster fullerenes, encloses not only metal atoms but also non-metals such as nitrides, oxides, sulphides, carbides, and related cyanides within the fullerene shell.[22, 28-37] Although among all fullerene based SMMs, Dy@$C_s$(6)-$C_{81}$N possess the largest $T_B$ value,[25] among cluster fullerenes with single magnetic centre, DyScS@$C_{82}$ exhibit the one of the best $T_B$ value of 7.3 K. Almost all reported EMFs exhibiting SMM characteristics within this range, suggesting a common spin-phonon relaxation mechanism in this class of molecules that may limit the achievement of higher $T_B$ values. In this manuscript, we investigated the spin-phonon relaxation processes in one such cluster fullerene DyScS@$C_{82}$ using state-of-the-art *ab initio* spin relaxation simulations. To the best of our knowledge, our study marks the first implementation of ab initio spin-phonon relaxation calculations in the context of fullerene-based SMMs. Our goal is to (i) elucidate the mechanism of spin-phonon relaxation, (ii) compute relaxation times for comparison with experimental data, (iii) analyse the vibrational modes responsible for spin-phonon relaxation, and (iv) provide general guidance on how to achieve larger $T_B$ values in this class of molecules.

In the original paper by Echegoyen and co-workers, [27] the synthesis and characterisation of two new dysprosium-containing mixed-metallic sulphide cluster fullerenes, DyScS@$C_s$(6)−$C_{82}$ and DyScS@$C_{3v}$(8)−$C_{82}$ were reported. These compounds were isolated and analysed using various techniques, including mass spectrometry, Vis-NIR spectroscopy, cyclic voltammetry, and single-crystal X-ray diffraction. For this manuscript, we chose DyScS@$C_s$(6)−$C_{82}$ for our calculations due to its higher $U_{eff}$ barrier and longer relaxation times in the absence of external magnetic fields compared to DyScS@$C_{3v}$(8)−$C_{82}$. In the remainder of the manuscript, we will refer to DyScS@$C_s$(6)-$C_{82}$ simply as DyScS@$C_{82}$.

**Computational Details**

To execute Cell and geometry optimizations, as well as simulations of Γ-point phonons, the CP2k program package[38-40] was used. The CP2k software uses a hybrid basis set technique called the Gaussian and Plane Wave Method (GPW)[39, 40], in which the electronic charge density is represented by an auxiliary plane-wave basis set and Kohn-Sham orbitals are enlarged using contracted Gaussian-type orbitals (GTOs). Utilising their relativistic norm-conserving pseudopotentials (Goedecker, Teter, and Hutter)[41] optimised for the PBE functional.[42] The Dy$^{III}$ ion was replaced by the Y$^{III}$ ion to avoid convergence issues with the DFT framework. The DZVP-MOLOPT-GTH basis set (valence double-zeta (ζ) plus polarisation, molecularly optimised, Goedecker-Teter-Hutter) for all atoms (H, C) and the DZVP-MOLOPT-SR-GTH basis set (valence double-zeta (ζ) plus polarisation, molecularly optimised, short range Goedecker-Teter-Hutter) for Y, as implemented in the CP2k, were used for all calculations.[40, 43] Furthermore, 1000 Ry was used as the energy cut-off for the plane wave basis set. For cell optimization, a very tight force convergence criterion of $10^{-8}$ au and an SCF convergence criterion of $10^{-8}$ au for energy were employed. The crystal field parameters for the geometries were computed using the ORCA 5.0 suite of software.[44] For the $Dy^{III}$ ions, magnetic properties were derived from CASSCF calculations with an active space of seven 4f orbitals containing nine electrons (9,7), considering all solutions with multiplicities of 6. The basis sets DKH-SVP for C atoms, and SARC-DKH-TZVP for Dy atoms[45], DKH-TZVP for Sc and S atoms were utilized.[46, 47] Also, we have performed CASSCF/RASSI-SO[48]/SINGLE_ANISO calculations using the Molcas8.2 package in order to compute the static relaxation mechanism in DyScS@$C_{82}$ molecule and the {Dy–S–Sc}$^{4+}$ fragment.[49, 50] We have employed the ANO-RCC-TZVP basis set[51] for $Dy^{III}$ ions and ANO-RCC-DZVP for rest of the atoms. We have used 21 sextets in CASSCF and RASSI-SO calculations as demonstrated earlier in this class of compounds.[5, 52, 53]

**Calculations of the Crystal Field and Spin-Phonon Coupling Coefficients**

The effective crystal-field Hamiltonian employed here correspond to

$$\hat{H}_{CF} = \sum_{l=2,4,6} \sum_{m=-l}^{l} B_m^l \hat{O}_m^l \quad (1)$$

The parameters $B_m^l$ are the CF Hamiltonian coefficients, and the operators $\hat{O}_m^l$ are tesseral functions of the total angular momentum operator $J$, with rank l and order m.[54] The lowest 2$J$+1 ab initio wave functions are matched one-to-one with the magnetic ground state of the ion to determine the CF Hamiltonian coefficients from first principles. To achieve this, the spin Hamiltonian $|\tilde{J},m_j\rangle$ is obtained by diagonalizing the operator $\hat{J}_z$ within the basis of the lowest 2$J$+1 ab initio wave functions.[55] Subsequently, the ab initio energy matrix is represented in this basis, and the parameters in Equation 1 are adjusted to reproduce the ab initio energy matrix elements.[55] These values enable the calculation of the magnetic anisotropy tensor across all temperatures and the magnetization in every spatial direction. The crystal field Hamiltonian, as described in equation 1, represents the energy levels of the ground-state electronic multiplet for the equilibrium geometry of the molecule. However, due to thermal energy, the molecular geometry is constantly fluctuating. These fluctuations lead to the modulation of spin properties, namely spin-phonon coupling, causing transitions between different electronic states until a thermal equilibrium is achieved between the electronic states and the lattice. Therefore, it is necessary to describe the spin, the phonons, and their interaction in detail in order to provide a quantum mechanical explanation of spin-dynamics. To derive the spin-phonon coupling coefficients ($\partial\hat{H}_s/\partial Q_\alpha$), $B_m^l$ that occur in equations (1) are numerically differentiated with respect to the atomic displacements described by $Q_\alpha$.

$$\left(\frac{\partial \hat{H}_S}{\partial Q_\alpha}\right) = \sum_{i=1}^{3N} \sqrt{\frac{\hbar}{2\omega_\alpha m_i}} L_{\alpha i} \left(\frac{\partial \hat{H}_S}{\partial X_i}\right) \quad (2)$$

where N is the number of atoms in the unit cell, $Q_\alpha$ is the displacement vector connected to the α-phonon, and $L_{\alpha i}$ and $\omega_\alpha$ are the eigenvectors of the Hessian matrix and the angular frequency of the phonon, respectively. By using numerical differentiation, the spin Hamiltonian's first-order derivatives ($\partial \hat{H}_S / \partial X_i$) with respect to the Cartesian degree of freedom $X_i$ are determined. Four samples are taken between ±0.1 Å for each molecular degree of freedom.

**Calculation of Spin-relaxation**

The total Hamiltonian operator is defined as:

$$\hat{H} = \hat{H}_S + \hat{H}_{Ph} + \hat{H}_{S-Ph} \qquad (3)$$

where the spin and phonon Hamiltonians are represented by the first two terms, respectively, and the coupling between these subsystems is represented by the third term. All-electron system's low-lying electronic states are described by the spin Hamiltonian. A straightforward sum of harmonic oscillators is used to simulate the phonon Hamiltonian. After obtaining the eigenstates, $|a\rangle$, and eigenvalues, $E_a$, of these operators, spin dynamics can be simulated by calculating the $W_{ab}$, or transition rate between distinct spin states.[56] In molecular Kramers systems with high magnetic anisotropy, one- and two-phonon processes contribute to spin relaxation. When one-phonon processes are taken into account, the transition rate between spin states, $\hat{W}_{ba}^{1-Ph}$ is

$$\hat{W}_{ba}^{1-Ph} = \frac{2\pi}{\hbar^2} \sum_\alpha \left| \left\langle b \left| \frac{\partial \hat{H}_0}{\partial Q_\alpha} \right| a \right\rangle \right|^2 G^{1-ph}(\omega_{ba}, \omega_\alpha) \qquad (4)$$

where the phrase ($\partial \hat{H}_0 / \partial Q_\alpha$) indicates the strength of the coupling between spin and the α-phonon $Q_\alpha$, and $\hbar\omega_{ba} = E_b - E_a$. The $G^{1-Ph}$ function reads

$$G^{1-Ph}(\omega, \omega_\alpha) = \delta(\omega - \omega_\alpha)\bar{n}_\alpha + \delta(\omega + \omega_\alpha)(\bar{n}_\alpha + 1) \qquad (5)$$

Where the Bose−Einstein distribution accounting for the thermal population of phonons is represented by $\bar{n}_\alpha = \left(\exp\left(\frac{\hbar\omega_\alpha}{k_B T}\right) - 1\right)^{-1}$, the Boltzmann constant is represented by $k_B$, and the Dirac δ functions enforce energy conservation during the absorption and emission of phonon by the spin system, respectively, here δ functions has been approximated using Gaussian smearing. The finite differentiation is used to calculate lattice harmonic frequencies ($\omega_\alpha/2\pi$) and normal modes ($Q_\alpha$) following geometry optimisation using DFT. The Orbach relaxation mechanism is explained by equation 2, where the spin moves from the fully polarised state $M_J = J$ to an excited state with an intermediate value of Ms before it can emit phonons again and return to $M_J = -J$. This process also occurs for states represented by the total angular momentum, *J*. A further route of relaxation to equilibrium is offered by two-phonon processes, leading to the Raman mechanism. Two-phonon spin-phonon transitions, or $\hat{W}_{ba}^{2-Ph}$ are modelled as

$$\hat{W}_{ba}^{2-Ph} = \frac{2\pi}{\hbar^2} \sum_{\alpha\beta} \left| T_{ba}^{\alpha\beta,+} + T_{ba}^{\beta\alpha,-} \right|^2 G^{2-ph}(\omega_{ba}, \omega_\alpha, \omega_\beta) \qquad (6)$$

Where $T_{\alpha\beta,\pm ba}$ is

$$T_{ba}^{\alpha\beta,\pm} = \sum_{c} \frac{\langle b|(\partial \hat{H}_S/\partial Q_\alpha)|c\rangle \langle c|(\partial \hat{H}_S/\partial Q_\beta)|a\rangle}{E_c - E_a \pm \hbar\omega_\beta} \quad (7)$$

$T_{\alpha\beta,\pm ba}$ entails the simultaneous contribution of all spin states $|c\rangle$, often known as a virtual state. All two-phonon processes, such as the absorption of two phonons, the emission of two phonons, or the absorption of one phonon and the emission of a second, are taken into account by $G_{2-ph}$. The latter method, which in this instance yields a $G_{2-ph}$ reading, determines the Raman relaxation rate.

$$G^{2-Ph}(\omega, \omega_\alpha, \omega_\beta) = \delta(\omega - \omega_\alpha + \omega_\beta)\bar{n}_\alpha(\bar{n}_\beta + 1) \quad (8)$$

After computing each matrix element $\hat{W}_{ba}^{n-Ph}$, $\tau^{-1}$ can be anticipated by diagonalizing $\hat{W}_{ba}^{n-Ph}$ and determining the smallest nonzero eigenvalue. The Orbach contribution to the relaxation rate, $\tau_{Orbach}^{-1}$, is obtained from the analysis of $\hat{W}^{1-Ph}$, while the Raman contribution, $\tau_{Raman}^{-1}$, is obtained from $\hat{W}^{2-Ph}$. Hence, the $\hat{W}$ calculation of the entire relaxation time is $\tau^{-1} = \tau_{Orbach}^{-1} + \tau_{Raman}^{-1}$. For Raman, we have rotated the molecule in the Eigen frame of g-tensors of the ground state KD and then applied a small magnetic field to lift the degeneracy.

## Results and Discussion

**Structure, Bonding and the Static Relaxation Mechanism in DyScS@C$_{82}$**

To begin with, we have optimised the geometry of DyScS@C$_{82}$ using periodic DFT methods (See computational details). The optimized structure shows a Dy-S bond distance of 2.514 Å (Figure 1(a)), which is within the range reported with the X-ray structure with a high degree of disorder. Additionally, we optimized the internal {Dy–S–Sc}$^{4+}$ fragment alone without the cage to evaluate the effect of the cage on the structure and magnetic properties. Without the cage restriction, the naked {Dy–S–Sc}$^{4+}$ fragment is almost linear with a Dy–S–Sc angle of 179.9°, and the Dy−S bond distance increases to 2.605 Å. We also performed Atoms in Molecules (AIM) analysis to investigate the nature of the bonding in both DyScS@C$_{82}$ and the naked {Dy–S–Sc}$^{4+}$ fragment within DFT formalism (see Table S2 for more details). The bonding between Dy$^{III}$ and the cage is found to be $\eta^1$ in nature, suggesting that Dy$^{III}$ is di-coordinated with one carbon atom of the cage. The |V(r)/G(r)| ratio of the S−Dy and Dy−C bonds is 1.3 and 1.2, respectively, indicating ionic bonds. In the naked {Dy–S–Sc}$^{4+}$ motif also, the Dy-S bond has a |V(r)/G(r)| ratio of 1.1, which again indicates the ionic nature of the bond.

Furthermore, we performed SA-CASSCF/RASSI/SINGLE_ANISO calculations using MOLCAS to evaluate the magnetic properties of DyScS@C$_{82}$ and {Dy–S–Sc}$^{4+}$ motif. For DySc@C$_{82}$, the splitting of eight Kramers doublets (KDs) is found to be 1011.5 cm$^{−1}$. The $g_{xx}$, $g_{yy}$, and $g_{zz}$ values of the ground KD are 0.002, 0.003, and 19.925 (Table S3), respectively, indicating strong Ising anisotropy. The angle between the ground KD $g_{zz}$ and the excited state $g_{zz}$ remains small up to the second excited KD (2.5°-5.9°) but increases to 20.9° in the third excited KD (Table S3).

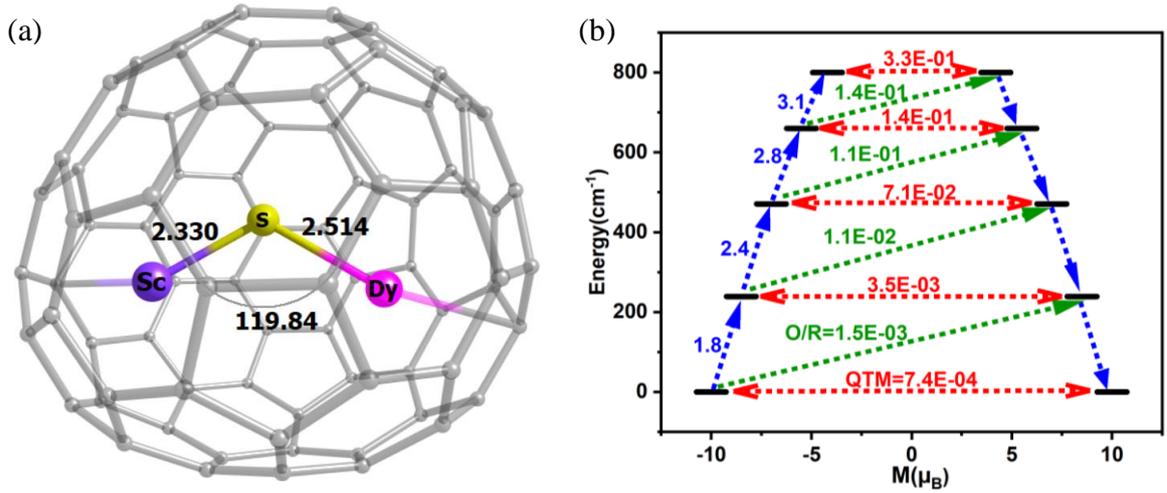

**Figure 1**: (a) DFT optimised structure of DyScS@C$_{82}$ (b) ab initio computed magnetic blockade diagram in DyScS@C$_{82}$. The thick black line indicates the KDs as a function of the computed magnetic moment. The green/blue arrows show the possible pathway through Orbach/Raman relaxation. The dotted red lines represent the presence of QTM/TA-QTM between the connecting pairs. The numbers provided at each arrow are the mean absolute values for the corresponding matrix element of the transition magnetic moment. colour code: Dy$^{III}$ -pink, S-yellow, C-grey and Sc-violet. bond lengths and bond angle are given in Å and °, respectively.

The third excited KD exhibits very strong transverse anisotropy ($g_{xx}$ = 0.641, $g_{yy}$ = 1.288, and $g_{zz}$ = 8.531), leading to relaxation from this state with an estimated theoretical barrier for magnetization reversal ($U_{cal}$) of 800.3 cm$^{-1}$ (Figure 1(b)). In contrast, the g tensors of the naked {Dy–S–Sc}$^{4+}$ fragment are purely Ising in nature up to the third excited state due to its completely linear geometry. Although the crystal field splitting (640.6 cm$^{-1}$) is lower in this case because Dy$^{III}$ is mono-coordinated, the angle between the ground KD $g_{zz}$ and the excited state $g_{zz}$ remains zero up to the fifth excited KD (Table S3). Relaxation occurs from the sixth excited KD, resulting in a relaxation barrier of 607.1 cm$^{-1}$ (Figure S1). Thus, despite relaxation happening via higher excited in the {Dy–S–Sc}$^{4+}$ motif, the $U_{cal}$ values are smaller due to longer Dy-S distance and absence of Dy-C interactions. This highlights the importance of cages in enhancing performance. Here it is noteworthy that the magnetic relaxation diagram computed using this approach is derived from transition matrix elements of the magnetic moment between opposite magnetization states (n+→n−) provides and it provides only a qualitative picture of the magnetic relaxation.

**Spin Phonon Relaxation in DyScS@C$_{82}$**

For first-order transitions, the rate at which spin states |a⟩ and |b⟩ change is given by an equation 5, first and second terms represent spin transitions caused by phonon absorption and emission, respectively. In perfectly axial Kramer systems without external time-reversal symmetry breaking interactions, direct $S_z$ = 15/2 → $S_z$ = -15/2 transitions are prohibited. Relaxation based on the given equation must occur through an excited state via phonon absorption. Whereas, second-order transitions involve two phonons simultaneously facilitating spin relaxation through an intermediate spin state |c⟩ within the S = 15/2 manifold. The population transfer rate between spin states |a⟩ and |b⟩ in this case is described by another equation (8). Here, $G^{2-Ph}$ accounts for energy conservation and phonon thermal population in two-phonon

interactions, similar to $G^{1-Ph}$ for single-phonon processes. This process is mediated by the excited spin state $|c\rangle$. The spin-relaxation time is calculated using the lattice force constants, spin-phonon coupling coefficients ($\partial\hat{H}_s/\partial Q_\alpha$), in Equations (4) and (6), as mentioned earlier. Figure 2 displays results from both first- and second-order perturbation theory, alongside the best fit to experimental data. First-order theory shows the expected exponential relationship between spin relaxation and temperature (T). Raman relaxation, however, The effective reversal barrier $U_{eff}$ and the pre-exponential factor ($\tau_0$) were determined by fitting the simulated Orbach data to the Arrhenius expression

$$\tau_{Orbach} = \tau'_0 e^{\frac{U_{eff}}{k_B T}} \quad (9)$$

In contrast, the results for Raman relaxation adhere to a more complex mathematical relationship. Recent literature has suggested that Raman relaxation is expected to follow a specific temperature-dependent law.

$$\tau^{-1}_{Raman} = \sum_i (\tau'_{0,i})^{-1} \frac{e^{\frac{W_{eff,i}}{k_B T}}}{\left(e^{\frac{W_{eff,i}}{k_B T}} - 1\right)^2} \quad (10)$$

But if there is only a single pair of phonons are involved, the Raman relaxations are also simplified to

$$\tau_{Raman} = \tau'_0 e^{\frac{W_{eff}}{k_B T}} \quad (11)$$

Based on Redfield equations, spin-phonon coupling coefficients are utilised to compute the spin-phonon relaxation time. One- and two-phonon processes have been simulated using second- and fourth-order density matrix time-dependent perturbation theory.[8, 56, 57] These simulations are conducted using the open-source programme MolForge, which can be downloaded from github.com/LunghiGroup/MolForge.

High temperatures see the dominance of the Orbach process, whereas the Raman process prevails at low temperatures. In Figure 2, by plotting $\ln(\tau)$ vs $1/T$, the slope of the curve at high temperatures gives the $U_{eff}$ value of 264 cm$^{-1}$, which also coincides with the energy of the first excited Kramers Doublet (275.2 cm$^{-1}$). The $U_{cal}$ value obtained from CASSCF calculations using the Molcas suite differs from this result because the previous method does not account for the effects of spin-phonon relaxation in its calculations.

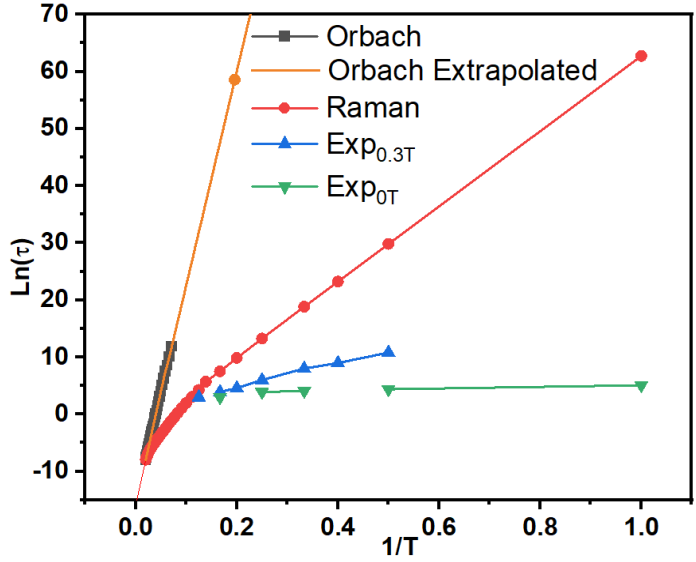

**Figure 2**: The relaxation times from the Orbach and Raman processes in comparison to the experimental relaxation times.

This aligns with the definition of the Orbach relaxation mechanism, which stipulates that phonons must be in resonance with the electronic states. However, the Orbach curve does not remain linear at low temperatures, indicating that other relaxation mechanisms, such as Raman relaxation, may become relevant. This is understandable because, at high temperatures, higher excited states are occupied, and phonons resonating with these states lead to relaxation. On the other hand, at low temperature a curve between ln(τ) vs 1/T for Raman relaxation, the $W_{eff}$ value comes out to be 49 cm$^{-1}$, which coincides with the first optical mode (45.7 cm$^{-1}$) at the Γ-point. Recent literature supports the idea that $W_{eff}$ should align with the lowest-energy phonons strongly coupled to the magnetic moment.[8, 58, 59] In previous papers papers some of us observed that this phonon typically corresponds to one of the lowest energy modes.[11] The relaxation times calculated using this method differ significantly from the experimental results, except at high temperatures, as shown in Figure 2. This difference can be explained by the omission of QTM in our approach, while the experimental studies are primarily influenced by QTM. Also it would be ideal to compare the experimental relaxation mechanism in a diluted sample for a more refined analysis, that data is currently unavailable. Additionally, the relaxation contribution from dipolar-mediated cross-relaxation, which is not included in our simulations, may account for some minor discrepancies between simulations and experiments. While incorporating anharmonic phonons could improve the accuracy of reproducing experimental results, recent literature suggests that even with their inclusion, achieving a perfect match remains challenging.

To gain deeper insights into the relaxation mechanisms of the DyScS@C$_{82}$ complex, we measured transition probabilities for Orbach and the Raman relaxation, as shown in Figure 3. Figure 3(a) presents computed transition rates for the Orbach mechanism at 10 K, while Figure 3(b) illustrates Raman relaxation probabilities at 5 K. It is noteworthy that these values represent the true *ab initio* computed spin-phonon transition rates used to calculate relaxation times, rather than the typical expected dipole moment values associated with such mechanisms. These diagrams suggest that the transition probability at the ground KD is very low, but it becomes higher in the excited states. Also, we observed that the intra-KD transition probability is higher in Raman relaxation in the ground state itself even than that of the third excited state in Orbach relaxation.

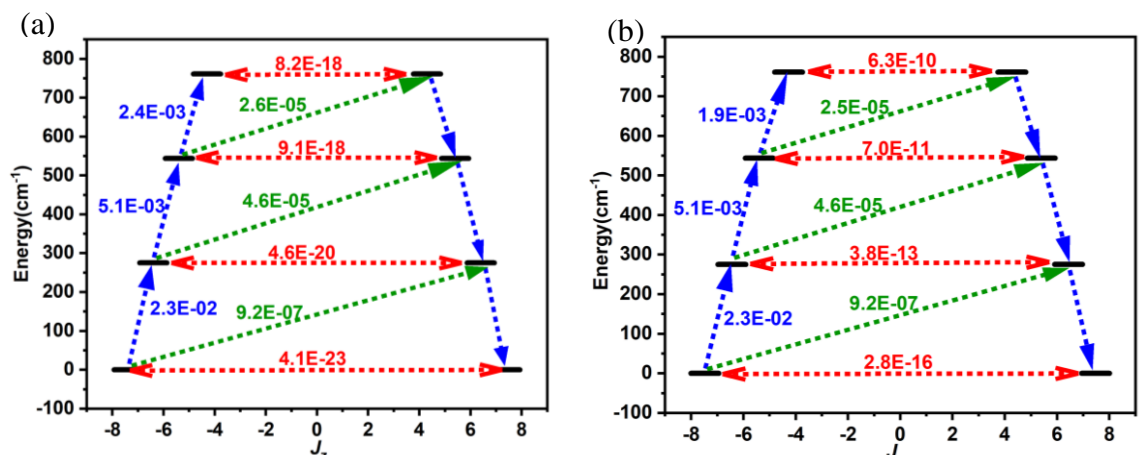

**Figure 3**: (a) Finite temperature Orbach transition rates and (b) Raman computed transition rates. The Orbach transition rates shows the matrix elements of $\hat{W}^{1-Ph}$ whereas the Raman relaxations shows the matrix element of $\hat{W}^{2-Ph}$. These transition rates are expressed in the basis of the eigenvectors of the ground KD's g-tensor. The x-axis displays the computed average magnetic moment for the first four KDs and their energy separation from the ground state.

At the fourth-order of perturbation theory, the relaxation process is dominated by the intra-ground state KD transition. A transition between two degenerate states necessitate of a simultaneous absorption and emission of two degenerate phonons in order for energy to be conserved. This is realized through a pair of phonons in the lowest part of the vibrational spectrum, as evinced by our previous discussion of the ln(τ) vs 1/T plot.

**Examination of Vibrational Density of States, Spin-Phonon Coupling Constants and the Molecular Vibrations**

After establishing the temperature dependence of the relaxation times, we focused on the spin-phonon coupling constants and the vibrational density of states (DOS) to decipher the origin of the Raman relaxation observed in the aforementioned section. Figure 4(a) illustrates the spin-phonon coupling coefficients as a function of frequencies along with the vibrational DOS. Figure 4(b) presents the detailed spin-phonon coupling coefficients alongside the energies of the eight Kramers doublets (KDs). We found no direct correlation between the vibrational DOS and spin-phonon coupling coefficients at low frequencies, though a partial correlation is evident at higher frequencies. Two regions exhibit very strong spin-phonon coupling: one near 50 cm$^{-1}$ (**vib1**, shown in Figure 5(a,b)) and another at 87 cm$^{-1}$ (**vib2**, shown in Figure 5(c,d)). Both modes occur at significantly lower frequencies than the first excited KD, indicating no possibility of Orbach relaxation for this molecule. Closer inspection of these modes revealed that they involve the movement of the Dy–S–Sc moiety inside the cage. Other spin-phonon coupling coefficients at higher frequencies are much smaller.

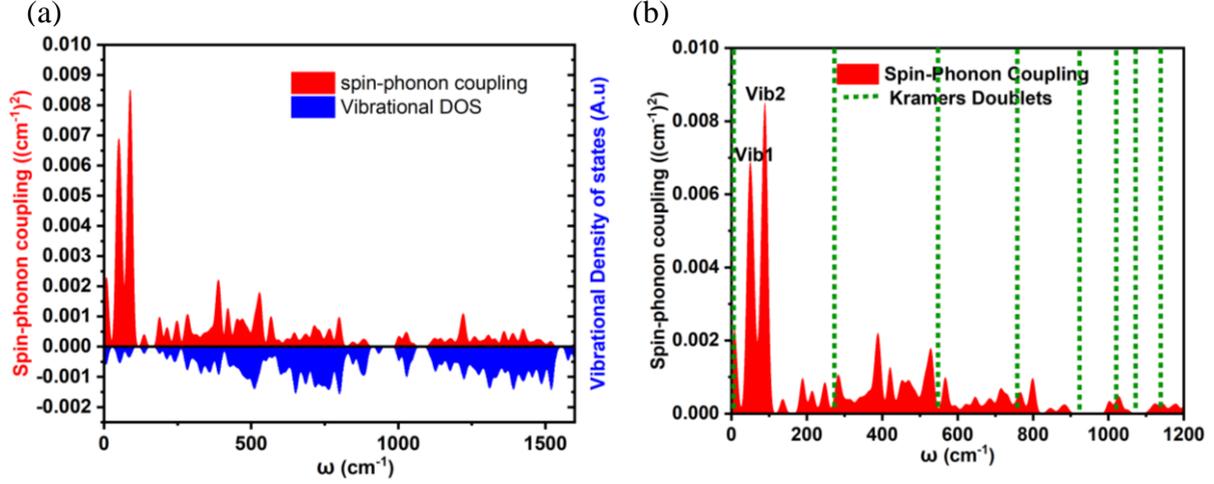

**Figure 4**: (a) spin-phonon coupling distribution and the vibrational density of states as functions of energy. A Gaussian smearing with σ = 1 cm$^{-1}$ and σ = 10 cm$^{-1}$ has been applied to the two functions, respectively. (b) close-up of the spin-phonon coupling distribution overlapped with the energies of the KDs.

As expected, since anisotropy should be primarily affected by the first coordination sphere, the highest spin-phonon coupling is observed with the movements of S, Sc, and Dy atoms inside the fullerene.

We will now discuss **vib1** and **vib2** in detail. Both modes are labelled in Figure 4(a) and are pictorially represented in Figure 5. These modes were selected based on their highest spin-phonon coupling coefficients and exhibit similar types of motions. **vib1** consists of the first two phonons at 45.7 cm$^{-1}$ and 54.0 cm$^{-1}$, involving two different vibrations. The first is a wagging-like vibration, where the Dy−S−Sc angle (and Sc−S, Dy−S bond distances) changes from 119.8° (2.331 Å, 2.514 Å) to 115.8° (2.395 Å, 2.550 Å). The second vibration is more like a scissoring motion, with Dy$^{III}$ moving more, changing the angle from 115.4° (2.333 Å, 2.532 Å) to 124.4° (2.328 Å, 2.518 Å). The **vib2** consists of two phonons at 79.8 cm$^{-1}$ and 90.4 cm$^{-1}$. The first phonon corresponds to a similar wagging mode as in vib1, where the bond angle and bond distances change from 117.7° (2.357 Å, 2.536 Å) to 119.8° (2.330 Å, 2.514 Å). The second phonon in **vib2** corresponds to a scissoring-like bending motion, with Sc$^{III}$ moving more, changing the angle from 117.1° (2.347 Å, 2.509 Å) to 122.7° (2.335 Å, 2.524 Å). As the {DySSc} unit inside the cage approaches the configuration of the naked {Dy–S–Sc}$^{4+}$ core, the relaxation mechanism is expected to shift. The {Dy–S–Sc}$^{4+}$ core, characterized by a more linear Dy−S−Sc bond angle and longer Dy−S distances, has a lower ground-state excited state gap. Inside the cage, the Dy−S distances elongate further, and the Dy−S−Sc bond angle widens closer towards the naked core, leading to the relaxation of magnetization.

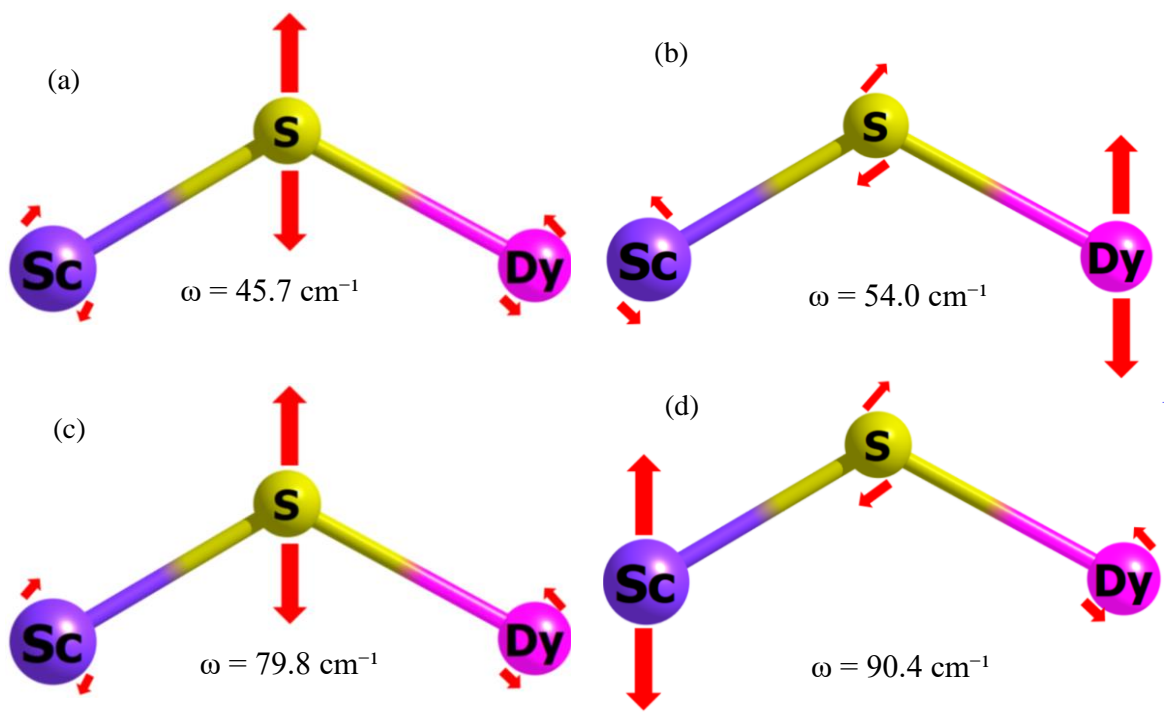

**Figure 5**: (a) First vibration of **vib1**, (b) second vibration of **vib1** (c) First vibration of **vib2** (d) second vibration of **vib2**. The red arrows indicate the movement, with the size of the arrow representing the extent of motion for that particular atom.

Additionally, it was observed that cage atoms begin to visibly vibrate only at the eighth phonon mode, which occurs at 210.9 cm$^{-1}$. This suggests that the cage indeed provides a largely rigid ligand environment that does not contribute to Raman relaxation.

**Discussion and Conclusions**

In this paper, first-principles simulations have demonstrated their capacity to accurately model spin relaxation processes. This approach enables the estimation of spin relaxation times without relying on experimental data, providing an unbiased means to benchmark our understanding of spin-phonon interactions and relaxation mechanisms. The field of molecular magnetism has made significant progress over the past three decades, with the discovery of compounds exhibiting remarkably $U_{eff}$ values and magnetic hysteresis at increasingly high temperatures. However, the field now faces a critical juncture, as traditional strategies for increasing the effective energy barrier in single-ion complexes may be approaching their practical limits. Future advancements in designing molecular compounds with extended spin lifetimes will likely stem from a more comprehensive understanding of the entire spin relaxation process. This necessitates a shift in focus beyond the singular pursuit of higher $U_{eff}$ values, encompassing a broader analysis of relaxation phenomena, including the often overlooked preexponential factor $\tau_0$. This computational approach enables researchers to directly interpret relaxation experiments without relying on phenomenological models. By providing a more fundamental and system-specific understanding, these simulations offer a powerful alternative to traditional interpretative frameworks, potentially leading to more accurate and insightful analyses of spin relaxation phenomena in molecular magnetic materials.

In addition to the low blocking temperature that limits single-molecule magnets, their practical use in information storage is hindered by the chemical instability of organometallic compounds, which prevents organized molecular assembly. Encapsulating lanthanide ions within chemically stable fullerene cages, forming EMFs, offers a solution. EMFs stabilize unique metal clusters with novel bonding configurations, enabling high-performance SMMs with greater structural diversity. EMFs, such as Dy@$C_{81}$N and $Dy_2$@$C_{79}$N, demonstrate enhanced magnetic properties, including high $T_B$ and strong magnetic coupling, and their chemical stability allows easy integration into spintronic devices. This structural and chemical robustness positions EMFs as a key avenue for advancing SMM applications and future synthetic developments in molecular magnetism. In this paper, for the first time, we have delved into the spin-phonon relaxation mechanisms within fullerene molecules, focusing on the high-performance SMM DyScS@$C_{82}$, which boasts a $T_B$ of 7.3K. The study of spin-phonon relaxation in DyScS@$C_{82}$ SMMs offers significant insights into how the interactions between molecular vibrations and electronic spin states influence the magnetic behaviour of these systems. The research highlights the importance of a molecular-level understanding of blocking temperatures in SMMs, which has long been a critical challenge in advancing practical applications of these molecules. The key-findings of this work are outlined below.

(i) **Role of cage in SMM performance:** The fullerene cage not only stabilizes the otherwise unstable fragment molecule but also modifies its geometry in a way that can enhance SMM performance. For example, in the case studied, the cage was found to enhance the crystal field splitting of the {DyScS} fragment compared to the naked {Dy–S–Sc}$^{4+}$ fragment (first to eighth KD state gap is 622.8 cm$^{-1}$ vs 1011.5 cm$^{-1}$ for DyScS@$C_{82}$). This indicates that the fullerene cage contributes to a much stronger crystal field. Both the Dy–C and shorter Dy–S bonds witnessed inside the cage contribute to this enhancement.

(ii) **Mode of Relaxation**: Analysis of relaxation time versus temperature graphs reveals that at low temperatures, *ab initio* computed Raman relaxation times closely align with experimentally determined values, indicating that the predominant relaxation mechanism is the Raman process. Furthermore, transition rates for Raman relaxation exceed those for the Orbach process. The highest spin-phonon coupling coefficients occur at low frequencies, significantly lower than the energy of the first Kramers doublet (KD), suggesting that the relaxation mechanism cannot be attributed to the Orbach process.

(iii) **Spin-Phonon Coupling Dynamics in DyScS@$C_{82}$:** At lower frequencies, we observed a pronounced phenomenon of spin-phonon coupling, where interactions between spin states and vibrational modes are notably strong. The lowest energy optical phonons arise from specific motions within the {Dy−S−Sc} moiety, exhibiting a robust coupling with the spin states of the system. It is noteworthy that vibrations originating from the fullerene cage occur at higher frequencies, beginning prominently at 210.9 cm$^{-1}$. These vibrations primarily represent the structural dynamics of the cage itself, with minimal interaction with the enclosed Dy−S−Sc moiety, thereby not leading to significant spin-phonon coupling. Thus, the cage clearly offers a significant advantage and does not contribute to spin-phonon relaxation.

(iv) **How Cluster Atom Vibrations Drive Relaxation**: We have identified four distinct low-lying phonons crucial to the relaxation dynamics of the system. These phonons involve specific bending motions that cause variations in the Dy−S−Sc bond angles, thereby initiating relaxation processes. The changes in Dy−S−Sc bond angles are known to influence the axial ligand field: an increase in bond angle reduces the axial ligand field strength, consequently promoting transverse anisotropy and facilitating relaxation pathways within the system. Unfortunately, these phonons predominantly operate at low frequencies, which poses a limitation on the system's ability to achieve high blocking temperatures. This characteristic underscores the importance of understanding and managing these low-frequency vibrational modes in the design and engineering of materials aimed at enhancing thermal stability and magnetic properties.

Our study reveals that while the fullerene cage improves stability and crystal field splitting of the {DyScS} fragment, it does not completely eliminate low-frequency vibrational modes within the cluster, which contribute to spin-phonon relaxation. These key low-lying phonons affect Dy−S−Sc bond angles, leading to transverse anisotropy and relaxation pathways. For enhanced performance in future EMFs, designs should focus on stronger Dy-ligand bonds, smaller cage sizes to reduce internal tumbling, and maintaining the cluster close to its equilibrium geometry. This is exemplified in the literature evidences for instance, DyLu$_2$N@I$_h$(7)-C$_{80}$, with a blocking temperature of 9 K[30], demonstrates stronger Dy-ligand interaction, while optimal combinations like in Dy@Cs(6)-C$_{81}$N[26], show fewer low-energy phonons coupled to spin states, leading to even higher blocking temperatures. This novel approach not only clarifies the spin-phonon mechanism for EMF-based SMMs but also provides a strategic pathway to enhance the performance of future EMF-based SMMs. Overall, this study advances the understanding of spin-phonon relaxation in SMMs and provides valuable guidelines for future molecular design aimed at increasing T$_B$ values. By focusing on the interplay between molecular structure, vibrational modes, and spin dynamics, it opens up new avenues for the development of high-performance SMMs, particularly in the context of fullerene-based systems where rigid molecular environments can stabilize low-coordinate lanthanide centers.

## Data Availability

The data supporting this article have been included as part of the Supplementary Information.

## Conflicts of Interest

There are no conflicts of interest to declare.

## Acknowledgement

TS is thankful to CSIR India for funding. GR would like to thank SERB (SB/SJF/2019-20/12; CRG/2022/001697) for funding. AL is thankful to the European Research Council (ERC) under the European Union's Horizon 2020 research and innovation programme (grant agreement no. 948493).

## References

1. F.-S. Guo, B. M. Day, Y.-C. Chen, M.-L. Tong, A. Mansikkamäki and R. A. Layfield, *Science*, 2018, **362**, 1400-1403.
2. C. A. Goodwin, F. Ortu, D. Reta, N. F. Chilton and D. P. Mills, *Nature*, 2017, **548**, 439-442.
3. S. K. Gupta, T. Rajeshkumar, G. Rajaraman and R. Murugavel, *Chemical Science*, 2016, **7**, 5181-5191.

4.   A. B. Canaj, S. Dey, E. R. Martí, C. Wilson, G. Rajaraman and M. Murrie, *Angewandte Chemie International Edition*, 2019, **58**, 14146-14151.
5.   T. Sharma, M. K. Singh, R. Gupta, M. Khatua and G. Rajaraman, *Chemical Science*, 2021, **12**, 11506-11514.
6.   C. A. Gould, K. R. McClain, D. Reta, J. G. Kragskow, D. A. Marchiori, E. Lachman, E.-S. Choi, J. G. Analytis, R. D. Britt and N. F. Chilton, *Science*, 2022, **375**, 198-202.
7.   A. Lunghi, F. Totti, R. Sessoli and S. Sanvito, *Nature communications*, 2017, **8**, 14620.
8.   M. Briganti, F. Santanni, L. Tesi, F. Totti, R. Sessoli and A. Lunghi, *Journal of the American Chemical Society*, 2021, **143**, 13633-13645.
9.   J. G. Kragskow, A. Mattioni, J. K. Staab, D. Reta, J. M. Skelton and N. F. Chilton, *Chemical Society Reviews*, 2023.
10.  M. Dolg, U. Wedig, H. Stoll and H. Preuss, *The Journal of chemical physics*, 1987, **86**, 866-872.
11.  S. Mondal and A. Lunghi, *Journal of the American Chemical Society*, 2022, **144**, 22965-22975.
12.  D. N. Woodruff, R. E. Winpenny and R. A. Layfield, *Chemical reviews*, 2013, **113**, 5110-5148.
13.  L. Ungur and L. F. Chibotaru, *Inorganic Chemistry*, 2016, **55**, 10043-10056.
14.  L. Ungur and L. F. Chibotaru, *Physical Chemistry Chemical Physics*, 2011, **13**, 20086-20090.
15.  J. D. Rinehart and J. R. Long, *Chemical Science*, 2011, **2**, 2078-2085.
16.  A. Ullah, J. Cerdá, J. J. Baldoví, S. A. Varganov, J. Aragó and A. Gaita-Ariño, *The journal of physical chemistry letters*, 2019, **10**, 7678-7683.
17.  A. Lunghi and S. Sanvito, *Science advances*, 2019, **5**, eaax7163.
18.  T. Sharma, A. Swain and G. Rajaraman, 2023, https://doi.org/10.26434/chemrxiv-2023-wpc4r.
19.  J. P. Durrant, J. Tang, A. Mansikkamäki and R. A. Layfield, *Chemical Communications*, 2020, **56**, 4708-4711.
20.  R. Nabi, R. K. Tiwari and G. Rajaraman, *Chemical Communications*, 2021, **57**, 11350-11353.
21.  G. Velkos, D. Krylov, K. Kirkpatrick, X. Liu, L. Spree, A. Wolter, B. Büchner, H. Dorn and A. Popov, *Chemical Communications*, 2018, **54**, 2902-2905.
22.  M. Nie, J. Xiong, C. Zhao, H. Meng, K. Zhang, Y. Han, J. Li, B. Wang, L. Feng and C. Wang, *Nano Research*, 2019, **12**, 1727-1731.
23.  L. Spree, C. Schlesier, A. Kostanyan, R. Westerström, T. Greber, B. Büchner, S. M. Avdoshenko and A. A. Popov, *Chemistry–A European Journal*, 2020, **26**, 2436-2449.
24.  Y. Wang, J. Xiong, J. Su, Z. Hu, F. Ma, R. Sun, X. Tan, H.-L. Sun, B.-W. Wang and Z. Shi, *Nanoscale*, 2020, **12**, 11130-11135.
25.  G. Velkos, W. Yang, Y.-R. Yao, S. M. Sudarkova, F. Liu, S. M. Avdoshenko, N. Chen and A. A. Popov, *Chemical Communications*, 2022, **58**, 7164-7167.
26.  Z. Hu, Y. Wang, A. Ullah, G. M. Gutiérrez-Finol, A. Bedoya-Pinto, P. Gargiani, D. Shi, S. Yang, Z. Shi and A. Gaita-Ariño, *Chem*, 2023, **9**, 3613-3622.
27.  W. Yang, G. Velkos, M. Rosenkranz, S. Schiemenz, F. Liu and A. A. Popov, *Advanced Science*, 2024, **11**, 2305190.
28.  C. Schlesier, L. Spree, A. Kostanyan, R. Westerström, A. Brandenburg, A. U. Wolter, S. Yang, T. Greber and A. A. Popov, *Chemical Communications*, 2018, **54**, 9730-9733.
29.  R. Westerström, J. Dreiser, C. Piamonteze, M. Muntwiler, S. Weyeneth, H. Brune, S. Rusponi, F. Nolting, A. Popov and S. Yang, *Journal of the American Chemical Society*, 2012, **134**, 9840-9843.

30. Y. Hao, G. Velkos, S. Schiemenz, M. Rosenkranz, Y. Wang, B. Büchner, S. M. Avdoshenko, A. A. Popov and F. Liu, *Inorganic Chemistry Frontiers*, 2023, **10**, 468-484.
31. D. Krylov, F. Liu, S. Avdoshenko, L. Spree, B. Weise, A. Waske, A. Wolter, B. Büchner and A. Popov, *Chemical Communications*, 2017, **53**, 7901-7904.
32. A. Kostanyan, C. Schlesier, R. Westerström, J. Dreiser, F. Fritz, B. Büchner, A. A. Popov, C. Piamonteze and T. Greber, *Physical Review B*, 2021, **103**, 014404.
33. G. Velkos, W. Yang, Y.-R. Yao, S. M. Sudarkova, X. Liu, B. Büchner, S. M. Avdoshenko, N. Chen and A. A. Popov, *Chemical science*, 2020, **11**, 4766-4772.
34. W. Yang, G. Velkos, F. Liu, S. M. Sudarkova, Y. Wang, J. Zhuang, H. Zhang, X. Li, X. Zhang and B. Büchner, *Advanced Science*, 2019, **6**, 1901352.
35. W. Yang, G. Velkos, S. Sudarkova, B. Büchner, S. M. Avdoshenko, F. Liu, A. A. Popov and N. Chen, *Inorganic Chemistry Frontiers*, 2022, **9**, 5805-5819.
36. A. Brandenburg, D. S. Krylov, A. Beger, A. U. Wolter, B. Büchner and A. A. Popov, *Chemical Communications*, 2018, **54**, 10683-10686.
37. W. Cai, J. D. Bocarsly, A. Gomez, R. J. L. Lee, A. Metta-Magaña, R. Seshadri and L. Echegoyen, *Chemical Science*, 2020, **11**, 13129-13136.
38. B. G. Lippert, J. H. PARRINELLO and MICHELE, *Molecular Physics*, 1997, **92**, 477-488.
39. J. VandeVondele, M. Krack, F. Mohamed, M. Parrinello, T. Chassaing and J. Hutter, *Computer Physics Communications*, 2005, **167**, 103-128.
40. T. D. Kühne, M. Iannuzzi, M. Del Ben, V. V. Rybkin, P. Seewald, F. Stein, T. Laino, R. Z. Khaliullin, O. Schütt and F. Schiffmann, *The Journal of Chemical Physics*, 2020, **152**.
41. G. Lippert, J. Hutter and M. Parrinello, *Theoretical Chemistry Accounts*, 1999, **103**, 124-140.
42. J. Paier, R. Hirschl, M. Marsman and G. Kresse, *The Journal of chemical physics*, 2005, **122**.
43. J. VandeVondele and J. Hutter, *The Journal of chemical physics*, 2007, **127**.
44. F. Neese, *Wiley Interdisciplinary Reviews: Computational Molecular Science*, 2012, **2**, 73-78.
45. D. A. Pantazis and F. Neese, *Journal of chemical theory and computation*, 2009, **5**, 2229-2238.
46. F. Weigend and R. Ahlrichs, *Physical Chemistry Chemical Physics*, 2005, **7**, 3297-3305.
47. F. Weigend, *Physical chemistry chemical physics*, 2006, **8**, 1057-1065.
48. P. Å. Malmqvist, B. O. Roos and B. Schimmelpfennig, *Chemical physics letters*, 2002, **357**, 230-240.
49. F. Aquilante, J. Autschbach, R. K. Carlson, L. F. Chibotaru, M. G. Delcey, L. De Vico, I. Fdez. Galván, N. Ferré, L. M. Frutos and L. Gagliardi, *Journal*, 2016.
50. L. F. Chibotaru and L. Ungur, *The Journal of chemical physics*, 2012, **137**.
51. B. r. O. Roos, R. Lindh, P.-Å. Malmqvist, V. Veryazov, P.-O. Widmark and A. C. Borin, *The Journal of Physical Chemistry A*, 2008, **112**, 11431-11435.
52. S. Dey, T. Sharma and G. Rajaraman, *Chemical Science*, 2024, **15**, 6465-6477.
53. S. Dey and G. Rajaraman, *Chemical Science*, 2021, **12**, 14207-14216.
54. A. Abragam and B. Bleaney, *Electron paramagnetic resonance of transition ions*, OUP Oxford, 2012.
55. L. Ungur and L. F. Chibotaru, *Chemistry–A European Journal*, 2017, **23**, 3708-3718.
56. A. Lunghi, *Science Advances*, 2022, **8**, eabn7880.
57. A. Lunghi and G. Rajaraman, *Chall. Adv. Comput. Chem. Phys*, 2023, 219-289.

58. A. Lunghi and S. Sanvito, *The Journal of Chemical Physics*, 2020, **153**.
59. L. Gu and R. Wu, *Physical Review Letters*, 2020, **125**, 117203.